\documentclass{article}
\usepackage{spconf,amsmath,graphicx}
\usepackage{amssymb}

\title{Particle Filter based Massive MIMO Channel Estimation}
\name{T. S. Anu$^\star$$^\dagger$, Tara Raveendran$^\star$$^\dagger$
}
\address{Author Affiliation(s)}
\address{$^\star$ Department of ECE, College of Engineering, Trivandrum, Thiruvananthapuram, India\\
	$^\dagger$ Affiliated to APJ Abdul Kalam Technological University, Kerala, India.
}
\begin{document}
%
\maketitle
\begin{abstract}
Massive multiple-input multiple-output (MIMO) communication systems have drawn significant interest recently in next-generation wireless communications. The use of a large number of antennas in massive MIMO makes the estimation of channel state information very challenging. Accurate channel state information is essential in capitalizing the advantages of the massive MIMO technology. This paper proposes the application of the Ensemble Square Root Filter (EnSRF) and a variant of EnSRF, namely a Particle wise Update version of the Ensemble Square Root Filter (PUEnSRF) to estimate the time-selective frequency-flat fading channel coefficients in the massive MIMO scenario. Simulation results clearly indicate the remarkably superior accuracy and filter convergence of PUEnSRF estimates as compared to the conventional particle filters.	
\end{abstract}
\section{Introduction}
\label{sec:intro}

One of the major challenges faced by the existing cellular networks is the increasing usage of smartphones, which in turn results in increased global mobile traffic.
The need for desired data rates is rapidly rising as a result of new applications like Augmented Reality (AR), Virtual Reality (VR), Internet of Things (IoT), etc. Currently existing 4G networks cannot potentially support the increasing data rates or a large number of connected devices envisaged in future networks \cite{1}.\\
Massive MIMO communication systems, explored in recent years, are a promising technology for the next generation of wireless communications. In massive MIMO, an array of antennas serves several individual users/terminals \cite{2}. Here, the number of users are fewer as compared to the base station (BS) antennas. The use of a large number of antennas in massive MIMO leads to significant spatial multiplexing gains, which increases the capacity of cellular networks by severalfold. Moreover, increasing the number of antennas results in a reduction in the radiated power (power scaling property of massive MIMO) and greater simplicity in signal processing. \\
Accurate channel state information is essential in capitalizing on the advantages of the massive MIMO technology. However, due to a large number of antennas, estimating channel state information is envisaged as one of the main challenges in such scenarios. By exploiting the channel reciprocity advantage of massive MIMO, the channel state information acquired in the uplink can be used in the downlink in time division duplex mode since the number of pilots required is proportional only to the number of users in the uplink channel \cite{3}.\\
The least-squares (LS) estimator is used for channel estimation in \cite{3}. However, the performance of the LS-based estimator is poor even though the estimation technique is simpler. In \cite{4}\cite{6}, the Minimum mean square error (MMSE) estimator is used for the channel estimation. MMSE can perform better than LS if an accurate channel correlation matrix is available. The matrix inversion process in MMSE makes the complexity of the estimator substantially larger than that of LS estimation, which in turn makes it unsuitable.\\
Large-scale fading which accounts for pathloss and shadowing effects cannot be disregarded in massive MIMO \cite{7}. The large-scale fading coefficient estimation error has a considerable impact on the system performance. The methods for estimating large-scale fading coefficients are described in \cite{8}\cite{9}.\\
For channel estimation based on pilots, the Kalman filter can improve the estimation accuracy \cite{10}. Kalman filter-based channel estimation \cite{11} can perform well even when a channel statistics mismatch exists between the acquired and actual channel state information.\\
Ensemble Square Root Filters (EnSRF) \cite{12} combines the statistical Monte Carlo sampling approach with the Kalman filter. For estimation, particle filters \cite{13} use a Bayesian approach. Particle Filters estimate the posterior density of the state space using a Bayesian recursive filter, which incorporates importance sampling. This technique imposes no constraints on the state space models or noise distribution. However, weight collapse that arises after a few iterations is the drawback of this technique. For this, resampling is done wherein new particles replace the particles with negligible weights in the vicinity of the particles with higher weights. The combination of sequential importance sampling with resampling used at each iteration is referred to as Sampling Importance Resampling (SIR), also known as Bootstrap filter \cite{14}.\\
The SIR approach is used for estimating the fast time-varying MIMO channel coefficients in \cite{15}. Two channel estimation techniques are proposed in \cite{16}, where one method uses a Gaussian particle filter, and the other uses an Auxiliary particle filter. An unscented particle filter is used for channel estimation in orthogonal frequency-division multiplexing MIMO system \cite{17}.\\
This paper explores the novel concept of applying the Ensemble Square Root Filter (EnSRF), and a variant of the EnSRF namely Particle wise Update version of the Ensemble Square Root Filter (PUEnSRF), to estimate the time-selective frequency-flat fading channel coefficients in massive MIMO scenario. In all the applications \cite{15}-\cite{17}, only small-scale fading has been considered whereas in this paper both large-scale and small-scale fading is considered.
Simulation results show that PUEnSRF gives better convergence and accuracy of estimates.

\section{System Model}
\label{sec:system_model}

The system model used here considers an uplink massive MIMO system with $N_r$ BS antennas and $N_t$ single antenna users. The channel gains are modeled as time-selective frequency-flat fading coefficients. The channel coefficient between $r$-th antenna of the BS and the $t$-th user in the $\tau^{th}$ transmission block is given by:
\begin{equation}
g_{rt,\tau} =  h_{rt,\tau} \;\sqrt{\beta _{t,\tau}} \;\;\;  ;\; 1\leq r \leq N_r ,\;\; 1\leq t \leq N_t
\end{equation}
where $\beta_{t,\tau} \in \mathbb{C}$, $h_{rt,\tau} \in \mathbb{C}$ denote the large-scale fading coefficient and small-scale fading coefficient, respectively.\\
The channel's time evolution can be modeled as a first-order autoregressive process \cite{18} as:
\begin{equation}
g_{rt,\tau}  = \alpha g_{rt,{\tau-1}} + \sqrt{1-|\alpha|^2} \gamma_{rt,\tau}
\end{equation}
where $\gamma_{rt,\tau} \in \mathbb{C}$ denotes the innovation noise. The temporal correlation coefficient $\alpha \in \mathbb{R}$ may be obtained from Jake’s model, $\alpha = J_0(2\pi f_D T_B)$ where $J_0(.)$ is the Bessel function of zeroth order, $f_D$ denotes maximum doppler frequency, and $T_B$ denotes block duration.\\
Channel coefficients between the user and base station in the $\tau^{th}$ transmission block may be expressed as
$\mathbf{G}_{\tau} \in \mathbb{C}^{N_r\times N_t}$ with $[\mathbf{G}]_{rt,\tau} = g_{rt,\tau}$.\\
The received signal (observation) in the $\tau^{th}$ transmission block, $\mathbf{Y}_{\tau} \in \mathbb{C}^{N_r\times N_t}$ is given as:
\begin{equation}
\mathbf{Y}_{\tau} = \sqrt{P_u}\mathbf{G}_{\tau} \mathbf{X}_{\tau} + \mathbf{N}_{\tau}
\end{equation}
where, $P_u$ is the uplink pilot transmit power, $\mathbf{X}_{\tau} \in \mathbb{C}^{N_t\times N_t}$ is the transmitted pilot symbols and $\mathbf{N}_{\tau} \in \mathbb{C}^{N_r\times N_t}$ is the additive white gaussian noise with $C\mathcal{N}(0,\sigma_n^2\mathbf{I})$.\\

\section{Channel Estimation Algorithm}
\label{sec:channel_est_Alg}
The algorithm to estimate the channel matrix $\mathbf{G}_{\tau}$ from the received signal $\mathbf{Y}_{\tau}$ and pilot matrix $\mathbf{X}_{\tau}$ using EnSRF and PUEnSRF is described below.
\subsection{Ensemble Square Root Filters (EnSRF)}
\label{sec:EnSRF}
EnSRF is a variant of Ensemble Kalman Filter (EnKF) \cite{12}. EnSRF avoids the Kalman filter Riccati update by approximating the error covariance using Monte Carlo simulation. This method is based on the presumption that prior knowledge of the matrix square root of forecast error covariance, henceforth referred to as forecast ensemble perturbation matrix, is available. Here the forecast and analysis ensemble perturbation matrices capture the information on error covariance. 
EnSRF has been used to estimate the channel coefficients at each transmission block, as follows.\\
An $N_r$-dimensional state space is represented for each transmission block by a t-member ensemble, $\{\mathbf{g}_{i,\tau} (i = 1,2,...t)\}$. The ensemble mean $\mathbf{\bar{g}}_{\tau}$ is represented by,
\begin{equation}
\mathbf{\bar{g}}_{\tau} = \frac{1}{t}\sum_{i=1}^{t}\mathbf{g}_{i,\tau}
\end{equation}
The $N_r$ x $N_t$ ensemble perturbation matrix $\mathbf{\tilde{G}}_{\tau}$ is represented as:
\begin{equation}
\mathbf{\tilde{G}}_{\tau} = \frac{1}{\sqrt{t-1}}[\mathbf{g}_{1,\tau}-\mathbf{\bar{g}}_{\tau} \;\; \mathbf{g}_{2,\tau}-\mathbf{\bar{g}}_{\tau} \; \; ... \;\; \mathbf{g}_{t,\tau}-\mathbf{\bar{g}}_{\tau}]
\end{equation}
The $N_r$ x $N_r$ ensemble covariance matrix $\mathbf{P}_{\tau}$ is defined as:
\begin{equation}
\mathbf{P}_{\tau} = \mathbf{\tilde{G}}_{\tau} \; \mathbf{\tilde{G}}_{\tau}^{\mathsf{T}}
\end{equation}
Let $\mathbf{Y}_{\tau}$ be the $N_r\times N_t$ observation matrix and $\mathbf{R}_{\tau}$ be the $N_r\times N_r$ observation error covariance matrix. 
The superscripts ‘$f$’ and ‘$a$’ denote the forecast(prior) and analysis(posterior) quantities, respectively.\\
The state space model of a dynamical system is often defined by process (state) and measurement (observation) equations.
A pseudo-dynamic approach is introduced for estimating the channel coefficients. The measurement equation is given by (3) and the process equation to estimate $\mathbf{g}_{\mathbf{i}+1,\tau}$ is given by:
\begin{equation}
\mathbf{g}_{i+1,\tau} = \mathbf{g}_{i,\tau} + \mathbf{\zeta}_{i+1,\tau}
\end{equation}
where $\mathbf{\zeta}_{i+1,\tau}$ is the noise vector.\\
Initially, an analysis ensemble $\{\mathbf{g}_{i,\tau}^{f}\}$, $i \in [1,2,...t]$ is constructed, whose mean indicates the estimate of the channel coefficient at $\tau^{th}$time instant. The forecast process ensemble, $\{\mathbf{g}_{i,\tau}^{f}\}$, $i \in [1,2,...t]$, is then obtained by applying process equation (7) to each member of the analysis ensemble. Using measurement equation (3), the forecast observation ensemble is then arrived at from the forecast process ensemble. Using equations (4) and (5), one can then find the ensemble mean vector and perturbation matrix for the forecast process ensemble and designate them as $\overline{\mathbf{g}}^f_{\tau}$ and $\mathbf{\tilde{G}}^f_{\tau}$, respectively. A similar sequence of steps is then adopted for arriving at the ensemble mean vector and perturbation matrix of the forecast observation ensemble and designate them as $\overline{\mathbf{y}}^f_{\tau}$ and $\mathbf{Y}^f_{\tau}$, respectively.\\
The analysis ensemble parameters are obtained as:
\begin{equation}
\mathbf{\overline{g}}^a_{\tau} = \mathbf{\overline{g}}^f_{\tau} + \mathbf{K}_{\tau}(\mathbf{y}_{\tau} - \mathbf{\overline{y}}^f_{\tau})
\end{equation}
\begin{equation}
\mathbf{P}^a_{\tau} = (\mathbf{\tilde{G}}^f_{\tau} - \mathbf{K}_{\tau}\mathbf{Y}^f_{\tau})(\mathbf{\tilde{G}}^f_{\tau})^\mathsf{T}
\end{equation}
\begin{equation}
\mathbf{K}_{\tau} = \mathbf{\tilde{G}}^f_{\tau} (\mathbf{Y}^f_{\tau})^\mathsf{T} \mathbf{D_{\tau}}^{-1}
\end{equation}
\begin{equation}
\mathbf{D}_{\tau} = \mathbf{Y}^f_{\tau} (\mathbf{Y}^f_{\tau})^\mathsf{T} + \mathbf{R}_{\tau}
\end{equation}

\subsection{A Particle wise Update version of the Ensemble Square Root Filter (PUEnSRF)}
\label{sec:PUEnSRF}
Unlike EnSRF, PUEnSRF does not rely on the Gaussian distribution assumption for the analysis ensemble in the update stage. Instead, in PUEnSRF \cite{19,20}, the ensemble as such, is propagated through the forecast and analysis stages. Hence, the analysis ensemble can be evaluated as:
\begin{equation}
\mathbf{\overline{g}}^a_{i,\tau} = \mathbf{\overline{g}}^f_{i,\tau} + \mathbf{K}_\tau (\mathbf{y}_\tau - \mathbf{\overline{y}}^f_{i,\tau})
\end{equation}
where $i$ represents the ensemble index.

\section{Numerical Illustrations and Results}

\begin{figure}[hbt!]
	\centering
	\includegraphics[width=0.4\textwidth]{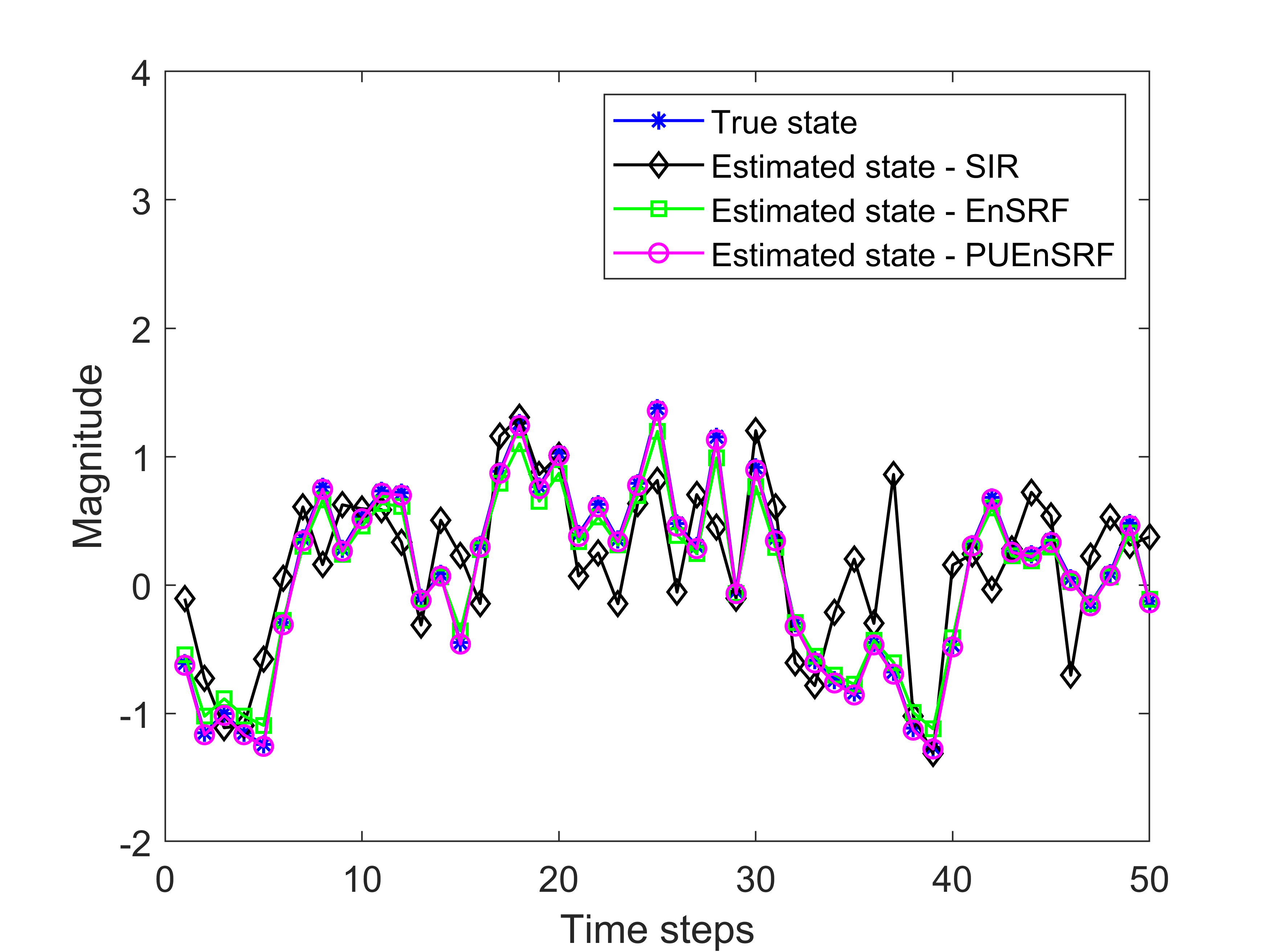} 
	\caption{Estimating the real part of the coefficient $g_{22}$ over the transmission blocks using SIR, EnSRF and PUEnSRF.}
	\label{fig:1}
\end{figure}

\begin{figure}[hbt!]
	\centering
	\includegraphics[width=0.4\textwidth]{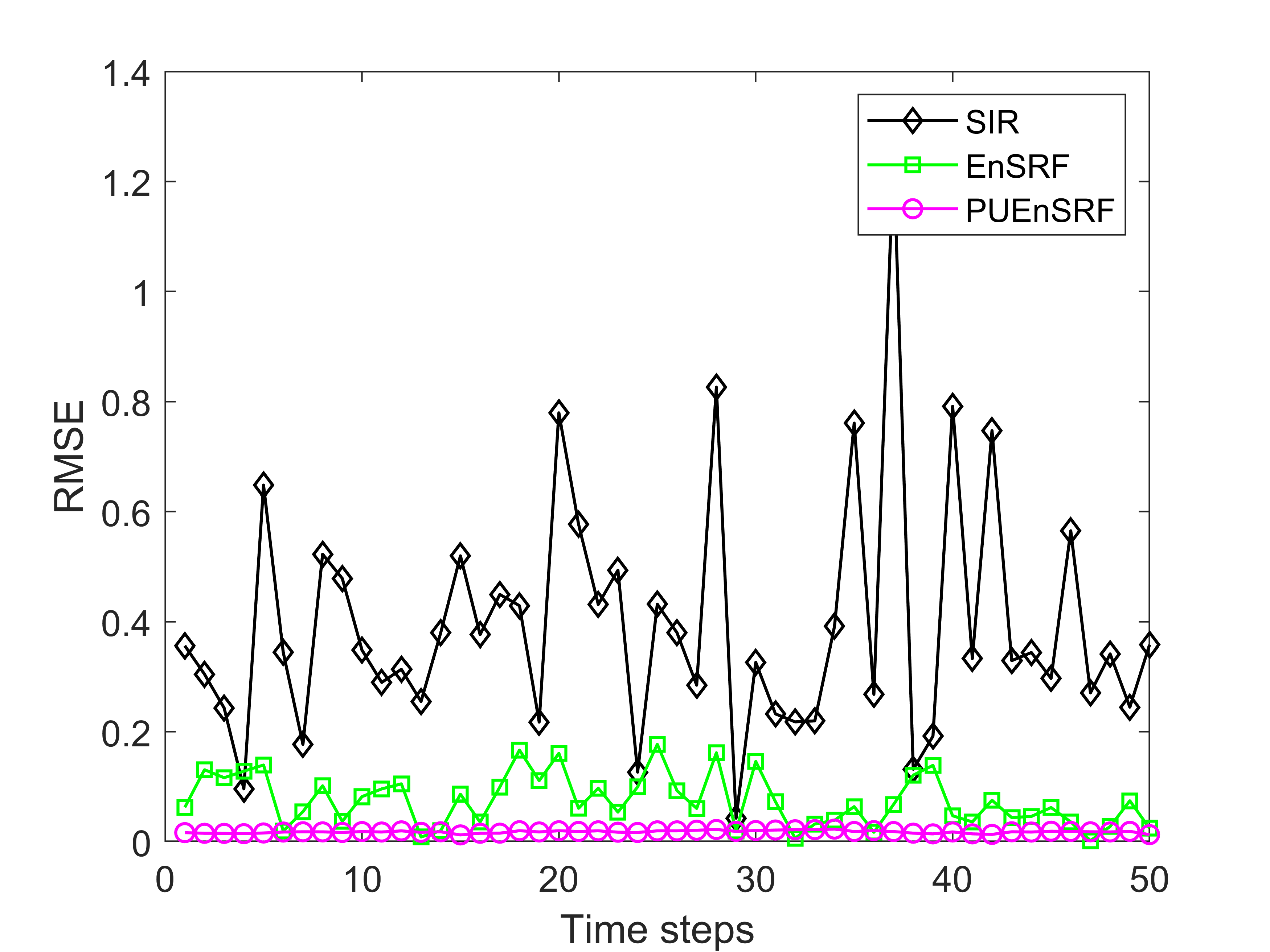} 
	\caption{RMSE of the real part of the coefficient $g_{22}$.}
	\label{fig:2}
\end{figure}

\begin{figure}[hbt!]
	\centering
	\includegraphics[width=0.4\textwidth]{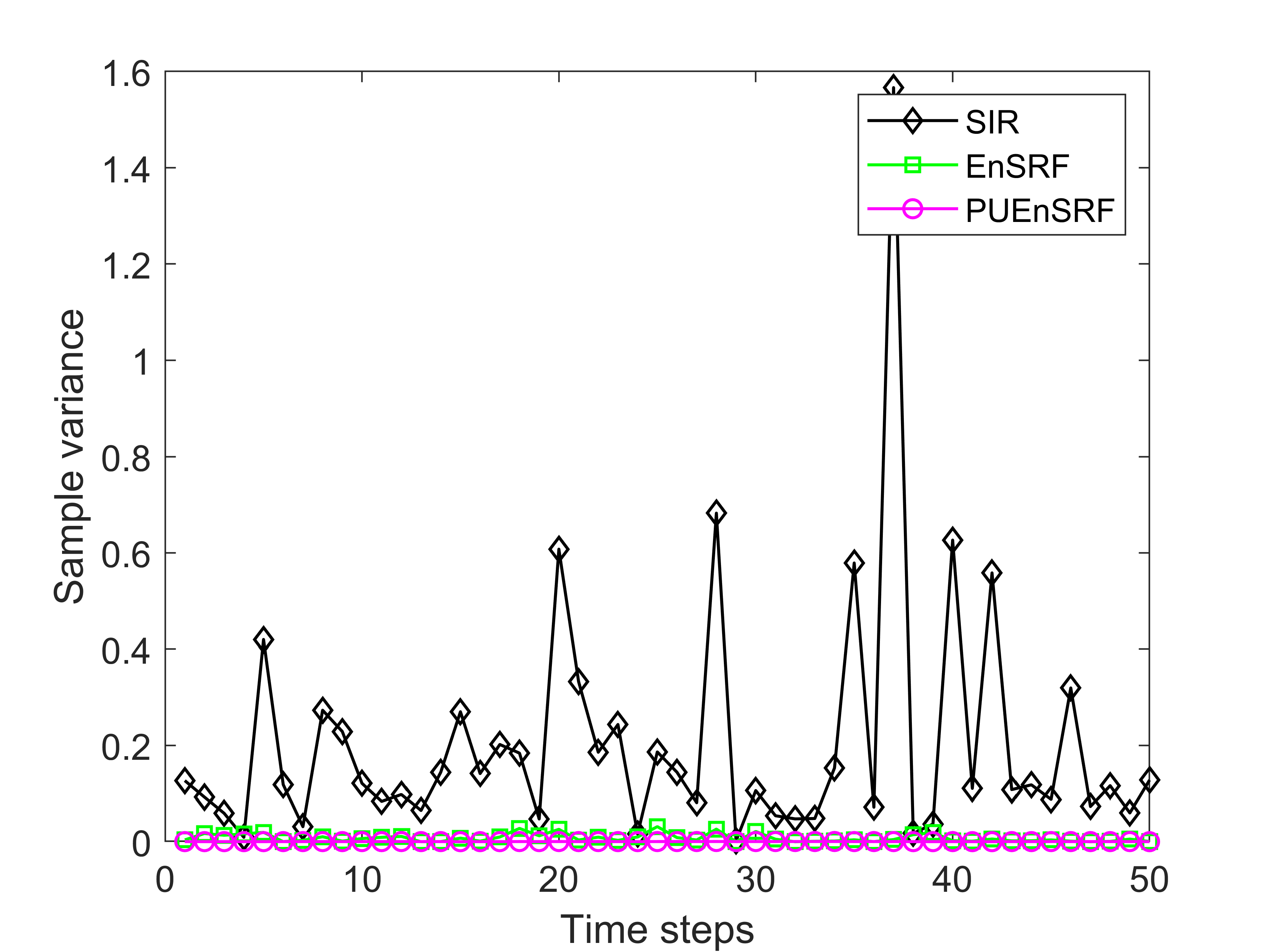} 
	\caption{Sample variance of the real part of the coefficient $g_{22}$.}
	\label{fig:3}
\end{figure}

\begin{figure}[hbt!]
	\centering
	\includegraphics[width=0.4\textwidth]{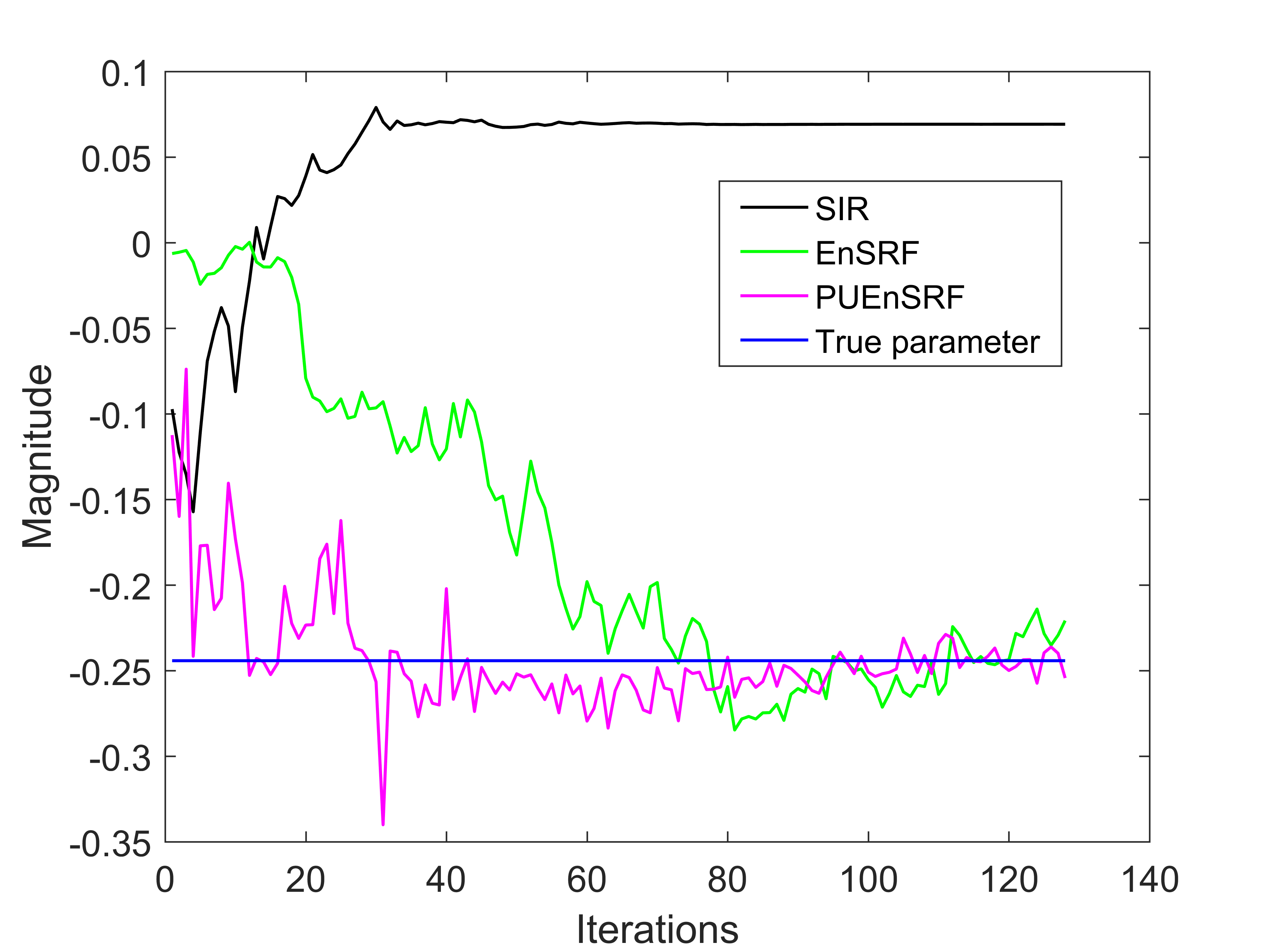} 
	\caption{Estimating the real part of the coefficient $g_{41} = -0.2442$.}
	\label{fig:4}
\end{figure}

\begin{figure}[hbt!]
	\centering
	\includegraphics[width=0.4\textwidth]{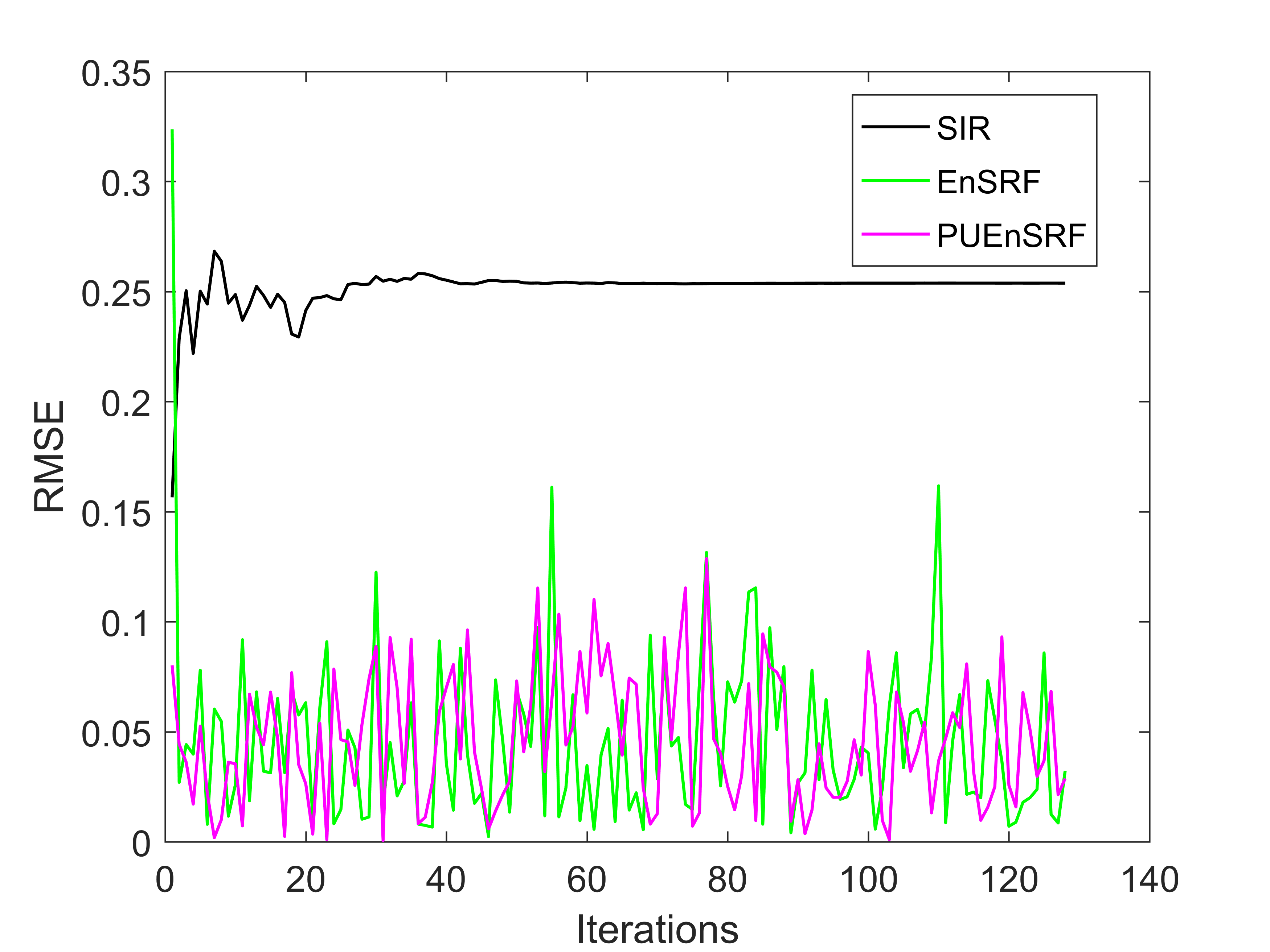} 
	\caption{RMSE of the real part of the coefficient $g_{41}$.}
	\label{fig:5}
\end{figure}

\begin{figure}[hbt!]
	\centering
	\includegraphics[width=0.4\textwidth]{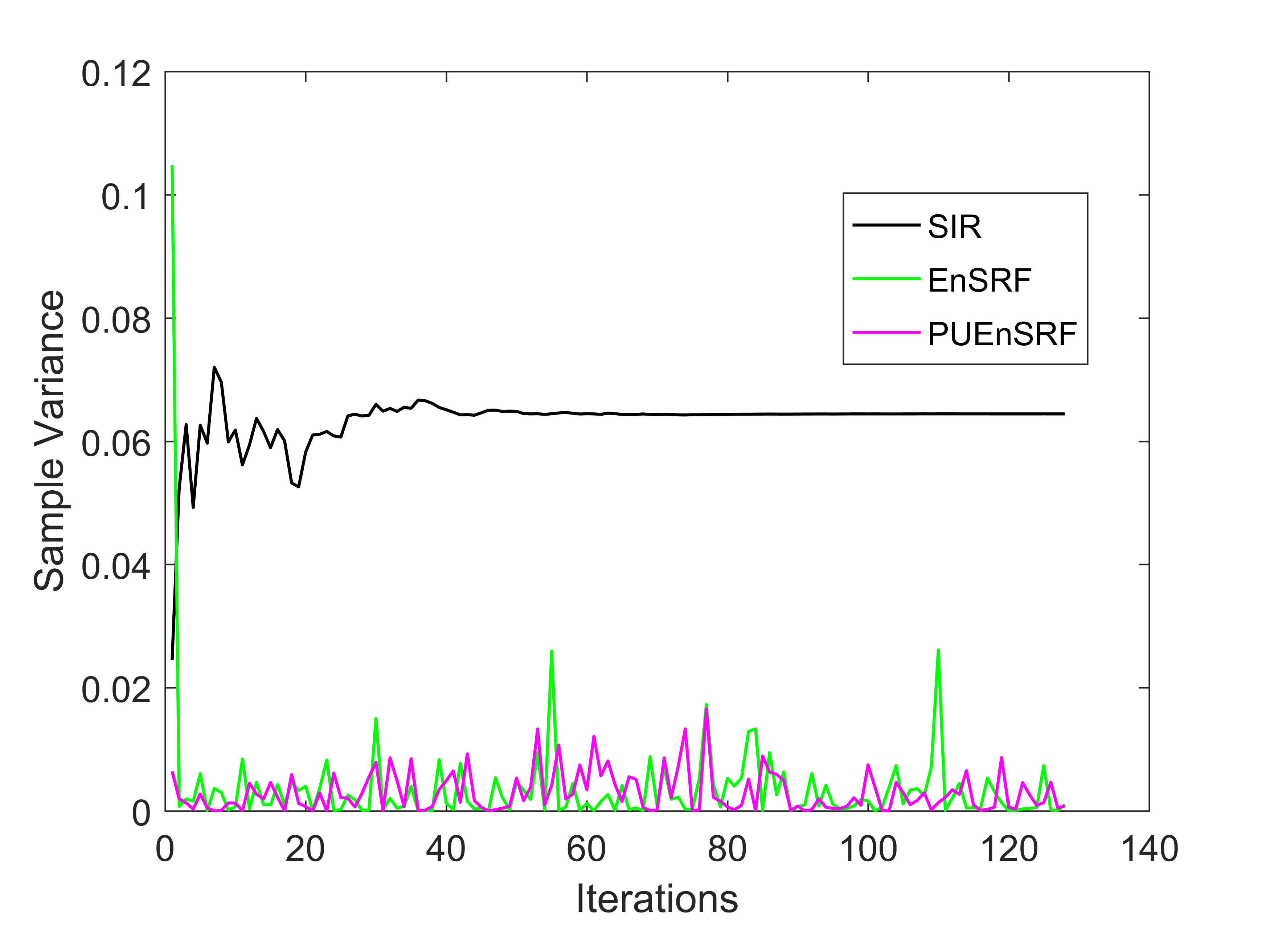} 
	\caption{Sample variance of the real part of the coefficient $g_{41}$.}
	\label{fig:6}
\end{figure}

The number of antennas at the transmitter ($N_t$) and receiver ($N_r$) are set as 128 and 512, respectively. The uplink transmit power is chosen to be 5 dB.
The large-scale fading in the $\tau^{th}$ transmission block is modeled as:
\begin{equation}
\beta_{t,\tau} = z_{t,\tau}/(d_{t,\tau})^{\gamma}
\end{equation}
where $z_{t,\tau}$ is log-normal random variable with standard deviation $\sigma_{shadow}$, $\gamma$ is the pathloss exponent, and $d_{t,\tau}$ is the distance between the $t$-th user and the BS.
Here $\sigma_{shadow}$ and $\gamma$ are set as 8dB and 3.8, respectively. The total number of transmission block is set as 50 (i.e. $1\leq \tau \leq 50$).\\
The massive MIMO system is deployed in the sub-6 GHz band with carrier frequency, $f_c$ = 2 GHz. The mobile velocity is considered to be $v$ = 20 km/hr which leads to a doppler shift of $f_D = f_c v/c$ = 37 Hz. $T_B$ = 2 ms is set as the block duration.
The temporal correlation coefficient $\alpha$ of the channel model in (2) is $\alpha = J_0(2\pi f_D T_B)$ = 0.95.\\
\textbf{Metrics used for evaluation:} The performance is evaluated using two metrics - Root Mean Squared Error (RMSE) and Sample variance. \\
RMSE is the square root of the average squared error between the actual value and estimated values.
\begin{equation}
\text{RMSE}(\tau) = \sqrt{\frac{1}{N_{MC}}\sum_{m=1}^{N_{MC}} (g_{rt,\tau} - \hat{g}_{rt,\tau})^2 }
\end{equation}
where $N_{MC}$ is the number of independent Monte Carlo runs and $\hat{g}_{rt,\tau}$ is the estimate of the channel coefficient between $r$-th antenna of BS and $t$-th user in the $\tau^{th}$ transmission block. $N_{MC}$ is chosen to be 50.\\
Sample variance is the average of the squared deviations from the sample mean. 
\begin{equation}
\text{Sample variance} (\tau) = \frac{1}{N_{MC}-1} \sum_{m=1}^{N_{MC}}(\hat{g}_{rt,\tau} -\bar{g_{\tau}})^2
\end{equation}\\
Fig. \ref{fig:1} shows the estimation of the real part of the coefficient $g_{22}$ over the transmission blocks using SIR, EnSRF, and PUEnSRF. From the figure, it can be seen that the PUESnRF can track the channel more accurately compared to EnSRF and SIR.
\\
Fig. \ref{fig:2} and Fig. \ref{fig:3} show RMSE and sample variance, respectively, of the estimation of the real part of the coefficient $g_{22}$ over the transmission blocks. From the figures, it can be seen that the PUEnSRF has a low RMSE and low sample variance compared to EnSRF and SIR.
\\
\\
Specifically, for any one of these transmission blocks, the RMSE for $k$-th iteration is defined as:
\begin{equation}
\text{RMSE}^{(k)} = \sqrt{\frac{1}{N_{MC}} \sum_{m=1}^{N_{MC}} (g_{rt,\tau} - \hat{g}_{rt,\tau}^{m (k)})^2 }
\end{equation}
where $\hat{g}_{rt}^{(k)}$ is the estimate of the channel coefficient between $r$-th antenna of BS and $t$-th user from the $m$-th run in the $k$-th iteration with the total number of iterations set as 128.\\
Sample variance for $k$-th iteration is defined as:
\begin{equation}
\text{Sample variance}^{(k)} = \frac{1}{N_{MC}-1} \sum_{m=1}^{N_{MC}}(\hat{g}_{rt,\tau}^ {m (k)}-\bar{g_{\tau}})^2
\end{equation}\\
Fig. \ref{fig:4} shows the estimation of the real part of the coefficient $g_{41}$ in the first transmission block ($\tau = 1$). From the figure, it can be seen that PUEnSRF converges faster than SIR and EnSRF.
\\
Fig. \ref{fig:5} and Fig. \ref{fig:6} show RMSE and sample variance, respectively, of the estimation of the real part of the coefficient $g_{41}$ in the first transmission block. From the figures, it can be seen that the PUEnSRF exhibits lower RMSE and sample variance compared to EnSRF and SIR.\\
\\
\\
\section{Conclusions}
A variant of EnSRF, namely PUEnSRF is explored for estimating the time-selective frequency-flat fading channel coefficients in massive MIMO scenarios. The simulation results show that PUEnSRF gives better convergence and accuracy of estimates when compared to SIR and EnSRF algorithms.

\bibliographystyle{IEEEbibAb}
\bibliography{refs1}


\end{document}